\documentclass[a4paper,12pt]{article}

\usepackage[dvips]{graphicx}
\usepackage{pstricks}
\usepackage{bm}

\usepackage{graphicx}

\newcommand{\be}{\begin{equation}}
\newcommand{\ee}{\end{equation}}

\newcommand{\beq}[1] {\begin{equation}\label{#1} }
\newcommand{\eeq} {\end{equation} }
\newcommand{\bea}[1]{\begin{eqnarray}\label{#1} }
\newcommand{\eea}{\end{eqnarray}}

\def\beqn{\begin{eqnarray}}
\def\eeqn{\end{eqnarray}}

\def\beq{\begin{equation}}
\def\eeq{\end{equation}}
\def\bea{\begin{equation}}
\def\eea{\end{equation}}
\def\pl{\partial}

\def\Ga{\Gamma}

\def\De{\Delta}

\def\Si{\Sigma}

\def\lam{\lambda}

\def\fr{\frac}
\def\la{\label}
\def\hs{\hspace}
\def\vs{\vspace}

\begin{document}
\vspace*{-0.2in}
\begin{flushright}
OSU-HEP-09-03\\
May 16, 2009\\
\end{flushright}

\vs{0.5cm}

\begin{center}
{\Large\bf New Mechanism for Neutrino Mass Generation and \\[0.1in]
Triply Charged Higgs Bosons at the LHC}\\
\end{center}

\vspace{0.5cm}
\begin{center}
{\large
{}~K.S. Babu\footnote{E-mail: babu@okstate.edu},{}~
S. Nandi\footnote{E-mail: s.nandi@okstate.edu}, and
{}~Zurab Tavartkiladze\footnote{E-mail: zurab.tavartkiladze@okstate.edu}
}
\vspace{0.5cm}

{\em Department of Physics \\
and Oklahoma Center for High Energy Physics\\
Oklahoma State University\\
Stillwater, OK 74078, USA }
\end{center}

\begin{abstract}

We propose a new mechanism for generating small neutrino masses which
predicts the relation $m_\nu \sim v^4/M^3$, where $v$ is the electroweak scale,
rather than the conventional seesaw
formula $m_\nu \sim v^2/M$.   Such a mass relation is obtained via effective
dimension seven operators $L L H H (H^\dagger H)/ M^3$, which arise
when an isospin $3/2$ Higgs multiplet $\Phi$ is introduced along with
iso-triplet leptons.  The masses of these particles are naturally in the TeV
scale. The neutral member of $\Phi$ acquires an induced vacuum
expectation value and generates neutrino masses, while its triply
charged partner provides the smoking gun signal of this scenario.
These triply charged bosons can be pair produced at the LHC and the
Tevatron, with $\Phi^{+++}$ decaying into $W^+\ell^{+}\ell^{+}$ or
$W^+ W ^+ W^+$, possibly with displaced vertices. The leptonic
decays of $\Phi^{+++}$ will help discriminate between normal and
inverted hierarchies of neutrino masses. This scenario also allows
for raising the standard Higgs boson mass to values in excess of 500
GeV.

\end{abstract}

\newpage

\section{Introduction}\label{sec:1}

The existence of small neutrino masses in the range $(10^{-2} -
10^{-1})$ eV has now been firmly established from a variety of
neutrino oscillation experiments and serves as the only direct
evidence for physics beyond the Standard Model. A question of
fundamental importance is how such tiny masses, many orders of
magnitude below their charged fermion counterparts, are generated.
An understanding of this question will also reveal if the neutrinos
are Dirac particles, very much like the charged leptons, or if they
are Majorana particles, very distinct from all other fermions.

The most compelling and popular explanation is the seesaw mechanism
\cite{seesaw} which generates neutrino masses via effective
dimension five operators $L L H H/M$, where $L = (\nu, ~e)$ is the
lepton doublet, $H = (H^+, ~ H^0)$ is the SM Higgs doublet, and $M$ is
the scale of new physics.  This leads to the mass relation for light
neutrinos $m_\nu \sim
v^2/M$, where $\left\langle H^0 \right\rangle = v$. These operators
can arise at tree level by integrating out heavy right--handed neutrinos
transforming as $(1,1,0)$ under $SU(3)_c \times SU(2)_L \times U(1)_Y$ (Type I
seesaw), or  $(1,3,2)$ Higgs bosons (Type II seesaw) \cite{TypeII},
or $(1,3,0)$ fermions (Type III seesaw) \cite{TypeIII}, all with
mass of order $M$. Neutrino oscillation data suggests $M \sim
10^{14}$ GeV, indicating that the associated new physics is likely
to be not within reach of colliders such as the Tevatron and the
LHC.  While there are several indirect benefits for the seesaw,
especially the one mediated by heavy right--handed neutrinos
(unification of all members of a family in $SO(10)$ grand unified
theory, leptogenesis), it is difficult to fathom a direct
verification of the mechanism in low energy experiments.  We feel
that it is important to explore alternative  mechanisms \cite{loop,
gn,zurab} which may be more directly tested  \cite{han,other}.

In this paper we propose a new mechanism for generating tiny
neutrino masses at tree level via effective dimension seven
operators $LL HH (H^\dagger H)/M^3$.  Such operators lead to
a new formula for neutrino masses, $m_\nu \sim v^4/M^3$, which is
distinct from the standard seesaw formula. Owing to the higher
dimensionality of these operators, neutrino mass generation will be
more readily accessible to collider experiments in this case.
Suppose that the $(1,1,0)$ and $(1,3,0)$ fermions as well as the
$(1,3,2)$ Higgs bosons are not present in the fundamental theory. In
this case $d=5$ neutrino mass operators will not be induced.  (Such
operators may still be induced by Planck scale physics, but this
will lead to extremely small neutrino masses of order $10^{-5}$ eV,
which are not relevant for neutrino oscillation data.) In such a
setup, the $d=7$ operators can be the source of neutrino masses. The
simplest realization of this mechanism  assumes the existence of an
isospin $3/2$ Higgs boson $\Phi = (\Phi^{+++}, \Phi^{++}, \Phi^+,
\Phi^0)$ and a pair of vector-like fermions transforming as $(1,3,2)
+(1,3,-2)$ under the SM gauge symmetry.  $\Phi^0$ acquires an
induced vacuum expectation value (VEV) via its interactions with the
SM Higgs doublet $H$, and induces small neutrino masses. Since
the mass of the $\Phi$ can naturally be in the TeV range, this new
mass generation mechanism has interesting implications for the
physics that will be explored at the LHC (and for smaller masses, at
the Tevatron as well). We have analyzed the most interesting signals of this
model, which occurs in the production and decay of triply charged
Higgs boson $\Phi^{+++}$, with the possibility of displaced
vertices.

\section{Model}

Our model is based on the SM symmetry group $SU(3)_c
\times SU(2)_L \times U(1)_Y$.  In addition to the usual fermions,
we introduce a pair of vector--like leptons, $\Sigma =
(\Sigma^{++}, \Sigma^+, \Sigma^0)$ and $ \overline{\Sigma} =
(\overline{\Sigma^0}, \Sigma^{-}, \Sigma^{--})$ transforming as
$(1,3,2)$ and $(1,3,-2)$ respectively under the gauge group.  The Higgs sector
consists of an isospin $3/2$ multiplet $\Phi=(\Phi^{+++},
\Phi^{++}, \Phi^{+}, \Phi^{0})$ (with $Y=3$), in addition to  the SM doublet $H
= (H^+, H^0)$.

\begin{figure}
\hspace{-3cm}
\begin{center}
\leavevmode
\leavevmode
\vspace{2.5cm} \includegraphics{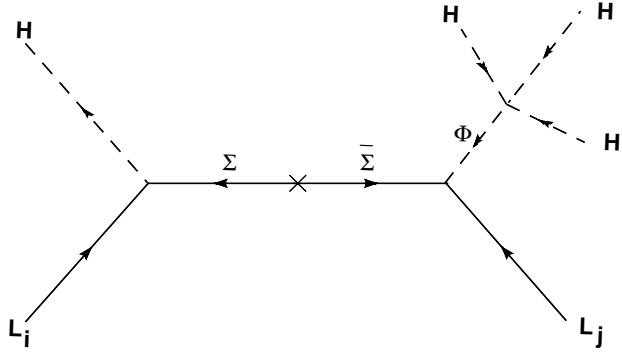}
\end{center}
\vs{2cm} \caption{Tree level diagram generating dimension 7 seesaw
operator for neutrino masses. } \label{fig1}
\end{figure}

Neutrino masses arise in the model from the renormalizable
Lagrangian
\beq {\cal L_{\nu-{\rm mass}}} = Y_iL_iH^*\Sigma +
\overline{Y}_iL_i\Phi \overline{\Sigma} +M_{\Sigma }\Sigma
\overline{\Sigma } + h.c., \la{Lsig} \eeq
where $Y_i,~\overline{Y}_i$
are Yukawa couplings and $i$ is the family index. Integrating out the $\Sigma , \overline{\Sigma
}$ fermions, one obtains an effective dimension five neutrino mass
operator \cite{zurab} \bea \la{neumass}
 {\cal L}_{\rm eff} = -{(Y_i \overline{Y}_j +  Y_j \overline{Y}_i) L_iL_j H^* \Phi \over M_\Sigma}+h.c.
\eea 
The tree level diagram generating this operator is shown in Fig. \ref{fig1}.
Analysis of the Higgs potential for $\Phi^0$ shows that it
acquires an induced VEV $\left\langle \Phi^0\right \rangle  \equiv
v_\Phi = -\lambda_5 v^3/M_\Phi^2$ where $\left\langle H^0
\right\rangle \equiv v$.  When this value is substituted  in Eq.
(\ref{neumass}), we obtain the neutrino masses to be $m^{\nu}_{ij} =
\lam_5(Y_i\overline{Y}_j+ Y_j \overline{Y}_i) v^4/(M_{\Sigma}
M^2_{\Phi})$.  This is the dimension seven mass generation
mechanism.  With $(Y_i, \overline{Y}_i, \lambda_5) \sim 10^{-3}$,
which are all in the domain of natural values, we obtain neutrino masses
in the $(10^{-2} - 10^{-1})$ eV range, consistent with neutrino
oscillation data, with $M_{\Si }$ and $M_{\Phi }$ in the TeV scale.
Eq. (\ref{neumass}) implies that one of the light neutrinos is
massless, which is consistent with current data. This feature arises
because we integrated a single pair of $\Sigma-\overline{\Sigma}$
fermions.  (One can readily add more than one pair of $\Sigma-\overline{\Sigma}$
states, in which case all neutrinos will acquire masses.)
Both normal hierarchy and inverted hierarchy of neutrino
masses can be realized with Eq. (\ref{neumass}). Note that our
neutrino mass relation  $m_{\nu} M^3 \sim v^{4}$ is distinct from
the  traditional seesaw relation $m_{\nu} M \sim v^2$ (modulo
dimensionless couplings).
While $d=5$ neutrino masses are not induced at tree level, they do arise at 1-loop in our model via diagrams which connect two of the $H$
legs in Fig. \ref{fig1} \cite{gavela}. We find these finite corrections to be $\De m_{\nu }/m_{\nu }\sim \fr{3}{64\pi^2}\fr{M^2}{v^2}$,
which is $\stackrel{<}{_\sim }1$ for $M\stackrel{<}{_\sim }2$~TeV.
In the SUSY version of our model, the loop diagrams will be further suppressed.

Now we assert the claim that $v_\Phi = -\lambda_5 v^3/M_\Phi^2$ used for neutrino mass estimates
by analyzing the Higgs potential.  It is given by
\begin{eqnarray}\label{pot}
V(H, \Phi) &=& \mu_H^2 H^\dagger H + \mu_\Phi^2 \Phi^\dagger \Phi + {\lambda_1 \over 2} (H^\dagger H )^2
+ {\lambda_2 \over 2}(\Phi^\dagger \Phi)^2 + \lambda_3 (H^\dagger H)(\Phi^\dagger \Phi) \nonumber \\
 &+& \lambda_4
(H^\dagger \tau_a H)(\Phi^\dagger T_a \Phi) + \{ \lambda_5 H^3 \Phi^* + h.c \}
\end{eqnarray}
where $\tau_a$ ($T_a$) are the generators of $SU(2)$ in the doublet (four-plet) representation.  We choose,
as usual in the SM, $\mu_H^2$ to be negative, so that the vacuum breaks electroweak symmetry. $\mu_\Phi^2$
will be chosen positive, yet due to the last term in Eq. (\ref{pot}), the neutral member of $\Phi$ will
acquire an induced VEV proportional to $v^3$.  Specifically, we have
\begin{equation}
v = (-\mu_H^2/\lambda_1)^{1/2},~~~~ v_\Phi = -\lambda_5 v^3/M_\Phi^2,
\end{equation}
where $M_\Phi^2 = \mu_\Phi^2 + \lambda_3 v^2 + {3 \over 4} \lambda_4
v^2$ is the mass of the neutral members in $\Phi^0$. (These
expressions ignore small corrections proportional to $v_\Phi^2$.)
The mass splittings between the members of $\Phi$ are given by
\begin{equation}
M_i^2 = M_\Phi^2 - q_i {\lambda_4 \over 4} v^2~
\end{equation}
where $q_i$ is the (non-negative) electric charge of the
respective $\Phi_i$ field.  We see that the mass splittings are
equally spaced and that there are two possible mass orderings.
For $\lambda_4$ positive, we have the ordering $M_{\Phi^{+++}} <
M_{\Phi^{++}}
 < M_{\Phi^{++}} < M_{\Phi^{0}}$, while for $\lambda_4$ negative,
 this ordering is reversed. We define a
(small)  splitting parameter  $\Delta M^2 \equiv (\lambda_4/4) v^2$
which is approximately equal to $(2 M_\Phi) (\Delta M)$.

The important phenomenological parameters of the model are
$v_{\Phi}$, $\Delta M$, $M_{\Phi}$ and $M_{\Sigma}$.   In this
paper, for simplicity, we shall assume that the triplet fermions
$\Sigma + \overline{\Sigma}$ have masses beyond the reach of LHC,
and focus on the signatures of the Higgs bosons from $\Phi$.  We
shall explore the entire range of 100 GeV to 1 TeV for $M_\Phi$. As
discussed, the VEV $v_\Phi$ can naturally be in the sub--MeV range for
$M_\Phi, M_\Sigma$ in the TeV range. $v_\Phi \neq 0$ modifies the
tree--level relation for the electroweak $\rho$ parameter, which is
now modified from 1 to $\rho \simeq  1-(6 v^2_{\Phi}/v^2)$.
Comparing with the experimental constraint of $1.0000 + 0.0011
(-0.0007)$ on $\rho$ \cite{Amsler:2008zzb}, we obtain, at 3 $\sigma$
level, $v_{\Phi} < 2.5$ GeV.  The mass splittings between the
components of $\Phi$ will induce an additional positive contribution
to $\rho$, given by $\Delta \rho \simeq (5 \alpha_2)/(6 \pi) (\Delta
M/m_W)^2$ \cite{veltman}. This sets an upper limit of $\Delta M < 38
$ GeV for the splitting parameter. There is also a theoretical lower
limit of $\Delta M > 1.4$ GeV, arising from (the finite parts) of
electroweak corrections \cite{strumia} which we shall comply with.
(This is actually a naturalness lower limit, since these corrections
are not finite, with the infinity absorbed in the renormalization of
the $\lambda_4$ coupling.) There are experimental lower limits on the
masses of $\Phi$, $M_\Phi > ~100$ GeV \cite{lep} for  a charged
$\Phi$ from LEP 2, and $M_\phi > ~120$ GeV  for a stable charged
$\Phi$ from the Tevatron \cite{tevatron}.

Production and decay of the $\Phi$ fields will depend on their gauge
interactions.  This is contained in the kinetic part of the
Lagrangian for $\Phi$, given as  ${\cal L}_{\rm kinetic}=(D^{\mu}
\Phi )^{\dag }(D_{\mu} \Phi)$ where \bea \la{dmuphi}
 D_{\mu} \Phi = \left(\pl_{\mu} -i g \vec{T} {\bf .}  \vec{W}_{\mu} - i g' {Y \over 2}B_{\mu}\right)\Phi ,
\eea
where $T_a$ are the $SU(2)$ generators for the isospin $(3/2)$
representations, and $Y$ is the hypercharge.

\section{Phenomenological Implications}

Since the mass scale for the new physics responsible for light
neutrino mass generation is naturally at the TeV, our model connects
neutrino physics to phenomena that can be observed at high energy
colliders, such as the LHC and the Tevatron. Specially, our model
has triply charged Higgs bosons which will have very distinctive
phenomenology, such as displaced vertices in the detectors, and
large number of $W$'s and/or charged multi-leptons in the final
states. Here we outline the decay characteristics of these bosons
and their production cross sections, and the ensuing final state
signals.

As noted earlier, the mass ordering inside the $\Phi$ multiplet
has two options. Here we consider the case in which $\Phi^{+++}$
is the lightest among the $\Phi$'s. This is the case in which
phenomenological implications of our model are most distinctive.
To study the signals of these bosons, we first consider the
various decay modes of $\Phi$'s and then the production cross
section at the Tevatron and the LHC.

\subsection{Decay of $\Phi$}
\begin{figure}
\hspace{-3cm}
\begin{center}
\leavevmode
\leavevmode
\vspace{2.5cm} \includegraphics{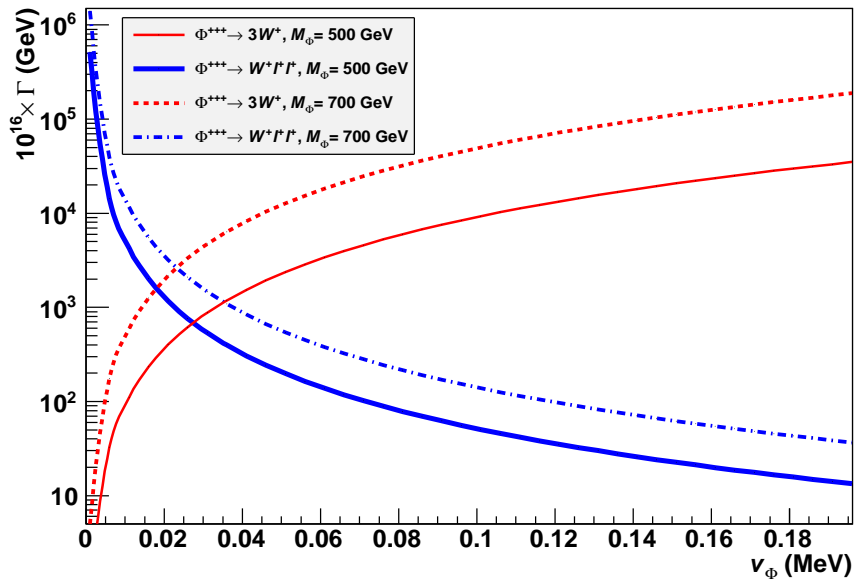}
\end{center}
\vs{4cm} \caption{Decay widths $\Ga (\Phi^{+++}\to 3W^{+})$ (red) and $\Ga (\Phi^{+++}\to W^{+}\ell^{+}\ell^{+})$ (blue) versus $v_{\Phi }$. Solid lines correspond to
$M_{\Phi^{+++}}=500$~GeV, and dashed curves to
$M_{\Phi^{+++}}=700$~GeV.} \label{fig2}
\end{figure}
$\Phi^{+++}$ has two principal decay modes: $\Phi^{+++} \rightarrow
W^+ W^+ W^+$ and $\Phi^{+++} \rightarrow W^+ \ell^+ \ell^+$. These
decays arise through the diagrams where $\Phi^{+++}$ emits a real
$W^+$ and  an off-shell $\Phi^{++}$ which subsequently decays to
either two real $W^+$, or two same sign charged leptons. The
relevant couplings are: $\Phi^{+++}\Phi^{--} W^- : \sqrt{3/2}g(p_1 -
p_2)_{\mu}$;\ $\Phi^{++} W^- W^- : \sqrt{3}g^2 v_{\Phi}$ ;
$\Phi^{++} \ell^-_i \ell^-_j : m^{\nu}_{ij}/(2\sqrt{3}v_{\Phi})$.  The
decay rates are found to be
\begin{equation}\label{width}
\Gamma(\Phi^{+++} \rightarrow 3W) ={3 g^6 \over 2048 \pi^3}
{v_\Phi^2 M_\Phi^5 \over m_W^6} I,~~~~ \Gamma(\Phi^{+++}
\rightarrow W^+ \ell^+ \ell^+) ={g^2 \over 6144 \pi^3} {M _\Phi
\sum_i m_i^2 \over v_\Phi^2} J
\end{equation}
where $I,J$ are dimensionless integrals.  In the limit where $M_\Phi
\gg m_W$, these integrals are approximately equal to one.  In Eq.
(\ref{width}), in the leptonic decay,  $m_i$ stand for the light
neutrino masses, and all flavors of leptons have been summed.  For
our numerical evaluation we have adopted the normal hierarchy of
neutrino masses with $m_3 = 0.05$ eV.  Results for the inverted
hierarchy spectrum can be obtained by multiplying the leptonic widths
by a factor of 2.

\begin{figure}
\hspace{-3cm}
\begin{center}
\leavevmode
\leavevmode
\vspace{2.5cm} \includegraphics{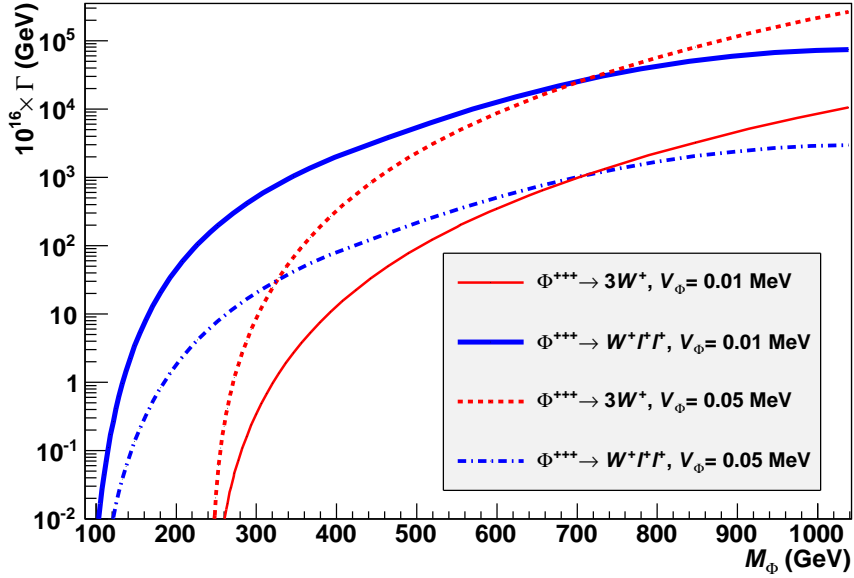}
\end{center}
\vs{4cm} \caption{Decay widths $\Ga (\Phi^{+++}\to 3W^{+})$ (red) and $\Ga (\Phi^{+++}\to W^{+}\ell^{+}\ell^{+})$ (blue) versus $M_{\Phi }$. Solid lines correspond to $v_{\Phi
}=0.01$~MeV, and dashed curves to $v_{\Phi }=0.05$~MeV.}
\label{fig3}
\end{figure}

The exact results for the partial decay widths in the $W^+W^+W^+$ mode and
 $W^+ \ell^+ \ell^+$ mode are shown in Fig. \ref{fig2} as a
function of $v_{\Phi}$ for $M_{\Phi} = 500$~GeV and $700$~GeV. The
same is shown in Fig. \ref{fig3} as a function of $M_{\Phi }$ for
$v_{\Phi }=0.01$~MeV and $0.05$~MeV. There are several interesting
characteristics for these decays. (i) Since the  $WWW$ mode is
proportional to $v^2_{\Phi}$, while the $W^+ \ell^+ \ell^+$ mode scales as
$1/v^2_{\Phi}$, the former is the dominant one for larger values of
$v_{\Phi}$, while the latter is dominant for smaller values of
$v_\Phi$. The two widths are equal at $v_{\Phi}\sim 0.03$ MeV for
$M_{\Phi}=500$ GeV. (ii) For widths $10^{-12}$ GeV or smaller,
$\Phi^{+++}$ will travel at least $0.2$ mm, and thus will produce
displaced vertex in the detector. For $M_{\Phi} = 500 $ GeV,  the
width for the $WWW$ mode equals $10^{-12}$ GeV for $v_{\Phi} = 0.1$
MeV, whereas for the $W^+ \ell^+ \ell^+$ mode, this happens for
$v_{\Phi} = 0.005$ MeV. Thus a $500$ GeV $\Phi^{+++}$ will produce
displaced vertex in the detector for a $v_{\Phi}$ range of $0.005 -
0.1$~MeV. For heavier (lighter) $\Phi^{+++}$, this range in
$v_{\Phi}$ is smaller (larger). A few hundred GeV $\Phi^{+++}$ can
travel as much as a meter in the detector before it decays. (iii)
 If $v_{\Phi}$ is much larger $0.1$ MeV, then it will decay
immediately inside the detector into $WWW$.

Let us now briefly comment on the decays of the heavier $\Phi$'s.
$\Phi^{++}$ has three principal decay modes: $\Phi^{++}\rightarrow
W^+W^+, \ell^+ \ell^+$,\ $\Phi^{+++} W^{-*}$, \ $\Phi^{+++} \pi^-$.
The first two decay rates depend on the value of $v_{\Phi}$, while
the remaining two depend crucially on the value of $\Delta M$. The
final states are $WW$, two same sign charged leptons, $WWWW^*$ or
$WWW \pi$. For the singly charged Higgs $\Phi^+$ (which mixes weakly
with the SM Higgs), decay modes are $\Phi^+ \rightarrow \Phi^{++}
W^{-*}, \Phi^{++} \pi^-$, with the subsequent decays of $\Phi^{++}$
as above. Thus the final states will be $WWW^*$, $WWWW^*W^*$ ,
$\ell^+\ell^+W^*$ , and so on. Finally for the neutral component,
$\Phi^0$ (which also mixes weakly with the SM Higgs), the final
state decay products can have as many as six $W$'s ($WWWW^*W^*W^*$),
or a combination of $W$'s and charged multi-leptons.

\subsection{Production of $\Phi$}

\begin{figure}
\hs{3cm}
\begin{center}
\leavevmode
\leavevmode
\vspace{0.5cm} \includegraphics{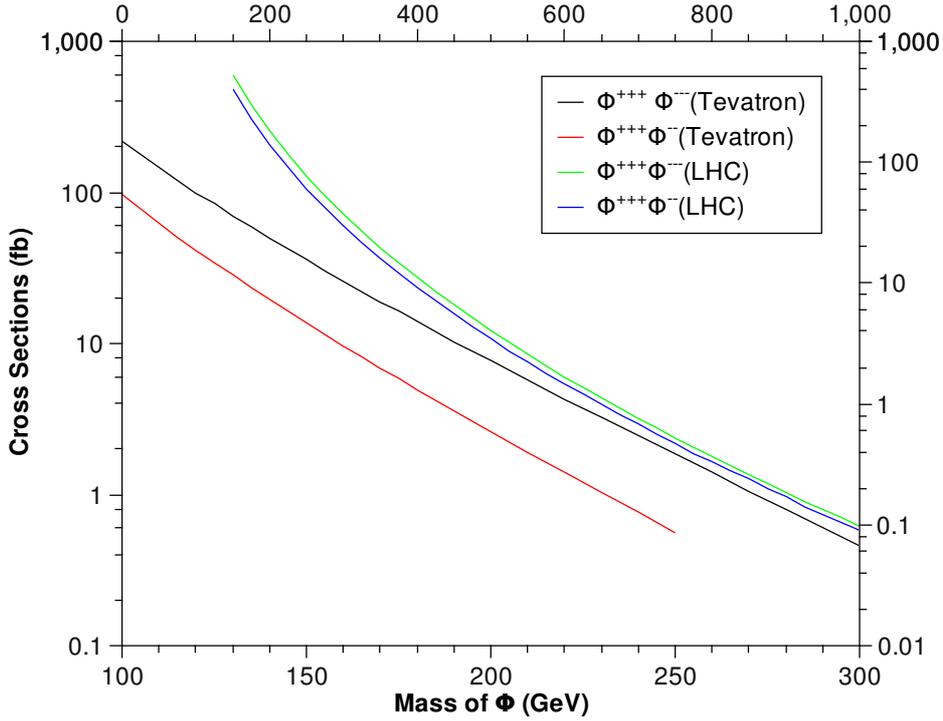}
\end{center}
\vs{7.4cm} \caption{Cross Sections (in fb) for $\Phi$ production vs
mass (in GeV) at the LHC
 (pp, $\sqrt{s} = 14$ TeV) and at the  Tevatron $(p \bar{p}, \sqrt{s} = 2$~TeV).
 The top horizontal and the right vertical axes are for the LHC, whereas the
 bottom horizontal and the left vertical axes are for the Tevatron.} \label{fig4}
\end{figure}

At the Tevatron and the LHC, $\Phi^{+++}\Phi^{--}$ and
$\Phi^{++}\Phi^{---}$ are pair produced via $s$-channel $W^+$ and
$W^-$ exchanges, while $\Phi^{+++}\Phi^{---}$ are pair produced
via the $s$-channel $\gamma$ and $Z$ exchanges. The $\Phi^{+++}
\Phi^{---} Z$ coupling is $-(3 e \cos
2{\theta_W}/\sin{2\theta_W})(p_1-p_2)_\mu$, while the $\phi^{+++} \phi^{--} W_\mu^-$
vertex is $\sqrt{3/2}~g ~(p_1-p_2)_\mu$.
The cross sections for $\Phi^{+++}\Phi^{--}$ production at the LHC
($pp$, $ \sqrt{s}=14$~TeV) and Tevatron $(p\bar{p},\sqrt{s}=2$~TeV)
are shown in Fig. \ref{fig4} as a function of the mass. We have
taken the masses of $\Phi^{+++}$ and $\Phi^{--}$ to be the same. (The
cross section for $\Phi^{++}\Phi^{---}$ production is approximately
a factor of five smaller at the LHC). At LHC, for a mass of 500 GeV,
$\Phi^{+++}\Phi^{--}$ production cross section is about 5 fb, while
it increase to about 60 fb for a mass of 250 GeV. For $v_{\Phi}$ in
the range of $0.005 - 0.1$ MeV, this will produce very distinctive
events with displaced vertices and five $W$ in the final state.  For
very small values of $v_{\Phi}$, the decay mode $\Phi^{+++}
\rightarrow W^+ \ell^+ \ell^+$ will dominate, with the final state being
$W^+ \ell^+ \ell^+ \ell^- \ell^-$, possibly with displaced vertex. Such events are
not expected in the SM, and will be a clear signal for new physics.
For the Tevatron, the cross sections are much smaller, however,
there may  still be such observable events. For example, for a $\Phi$
mass of 200 GeV, the cross section is about $2.6$ fb corresponding
to such displaced vertex events.

The cross sections for the $\Phi^{+++}\Phi^{---}$ production are
also shown in Fig.\ref{fig4}. At the LHC, for a mass of 500 GeV,
$\Phi^{+++} \Phi^{---}$ production cross section is about 4 fb,
while it increases to about 77 fb for a mass of 250 GeV. The final
states from these processes will produce $6 W$ or $2 W$ plus four
charged leptons. If $v_{\Phi}$ is $\leq 0.005$
MeV, the final state with 2$W$ plus four charged leptons will
dominate. The SM background for such events will be negligible. For
$v_{\Phi}$ in the range of $\sim 0.005 - 0.1$ MeV, both $6W$, and $2W$
plus 4 charged lepton final states will compete. However, both final
states will have displaced vertices very distinctive of new physics.
For $v_{\Phi}> 0.1$ MeV, 6W ($W^{+} W^{+ }W^+ W^- W^- W^-)$ final
state will dominate. Leptonic decays of any two same-sign $W$'s will produce same sign charged
dileptons with high $p_T$, with the final state having 8 high $p_T$ jets in addition
(from the decays of the other four $W$'s).
This will serve to reduce the SM
background severely, making the signal observable above the
background. Thus our model can be tested in the entire $v_{\Phi}$
range.
It is also possible that the lifetimes are so long that
$\Phi^{+++}$, $\Phi^{---}$ escape the detectors. This will produce
two tracks in the detectors characteristic of heavy triply charged
particles. If $v_{\Phi}$ is much larger, there will be no
displaced vertex, but the final states with multi-$W$ and
multi-leptons will be very distinctive. The cross sections for the
$\Phi^{+++}\Phi^{---}$ production for the Tevatron are also shown
in Fig.\ref{fig4}. These are somewhat larger than the
$\Phi^{+++}\Phi^{--}$ production cross section. Both the D0 and the CDF
experiments have looked for long-lived charged massive
particles which escape their detectors. Using the CDF upper limit on
the cross sections against the masses, we get a lower limit on such
a long-lived $\Phi^{+++}$ mass of $\sim 120$ GeV \cite{tevatron}.

We now briefly discuss the production of some of the heavier
states, and the ensuing final state signals. The cross section for
$\Phi^{++}\Phi^{-}$, for the same mass, is larger by a factor
$(4/3)$ compared to that of $\Phi^{+++}\Phi^{--}$ because of
slightly larger coupling, whereas the $\Phi^{++}\Phi^{--}$ pair
production cross section is significantly smaller than that for
$\Phi^{+++}\Phi^{---}$. However, if the masses are small enough so
that they are significantly produced, the final states are very
distinctive from their chain decays. For example, pair
production and subsequent decays lead to $\Phi^{+++}\Phi^{---}
\rightarrow 6W, \Phi^{++}\Phi^{--} \rightarrow 8W,~
\Phi^{+}\Phi^{-} \rightarrow 10W$, and $\Phi^{0} A^{0} \rightarrow
12W$ ($A^0$ being the neutral pseudoscalar), where some of the $W$'s
are off-shell. These are events with high charged lepton, or lepton
plus jet multiplicity,  all with high $p_T$, and are not expected in the SM.

\subsection{Other Implications}

Our model has interesting effects for the SM Higgs signals. The
isospin 3/2 Higgs multiplet $\Phi$ with a tiny VEV essentially behaves
like the inert Higgs \cite{Barbieri:2006dq}. SM Higgs boson ($H$)
mass can easily be raised to the range 400-600 GeV in our model. The
positive correction to the $\rho$ parameter proportional to $\Delta
M$, along with small corrections to the $S$ parameter,
weakens the usual bound of  185 GeV on the $H$ mass.  In this
case, there is an interesting possibility, which further enhances
the $\Phi$ production rate. $H$ can decay into a pair of triply
charged bosons (for $M_{\Phi}$ of order 200 GeV). The $H$ production
cross section at LHC via gluon fusion for $H$ mass of 500 GeV is
about 5 pb. With 1 inverse fb of data, there will be 5000 Higgs
events. $H$ will mostly decay, as in the SM, to 2 $W$'s. But the
branching ratio into triple charged Higgs pair is of order 10-20
percent (depending on the actual values of the quartic couplings
$\lambda_3$ and $\lambda_4$), implying 500-1000 events with possibly displaced vertices.
This will be a very unusual signal for the SM Higgs boson.

Our model also makes very interesting connections between the
neutrino mass hierarchy and collider physics. If the mass of
the $\Phi^{+++} <3 M_W$, then the decay mode $\Phi^{+++}
\rightarrow W^+\ell^+\ell^+$ will dominate, leading to final states
with $ee$, $e\mu$ or $\mu\mu$ (along with  $\tau$'s). The
dominance of $\mu\mu$ events will indicate normal hierarchy, while
that of $e\mu$ ($ee$) will indicate inverted hierarchy corresponding to
relative CP parity of the two heavier states being odd (even).  Since CP
symmetry is broken by the Majorana phases, these decay modes can be used to
measure these phases (see the first paper of Ref. \cite{han}).  Thus high
energy collider data will serve to distinguish between the models
of neutrino masses and mixings.

\section{Conclusion and Outlook}

We have presented a new mechanism for the generation of
neutrino masses via dimension seven effective operators $L L HH
(H^\dagger H)/M^3$. This leads to a new formula for light
neutrino masses, $m_\nu \sim v^4/M^3$, which is distinct from the
seesaw formula $m_\nu \sim v^2/M$.  The scale of new physics can quite naturally
be at the TeV scale.  This microscopic theory that induces the
$d=7$ operator has an isospin 3/2 Higgs multiplet containing
triply charged Higgs bosons.  We have analyzed the signatures of
these bosons at the LHC and the Tevatron.  Multi-lepton and multi
$W$ final states, with the possibility of displaced vertices,
should facilitate clean observation of these bosons.  The leptonic
decays of these particles carry information on the nature of
neutrino mass hierarchy, which can be directly probed at
colliders.

On a more
theoretical side, the model presented here can be extrapolated
from the weak scale all the way to the Planck scale without any
breakdown of  perturbation theory.  The isospin $3/2$ Higgs
representation can be readily embedded into the 35-dimensional
representation of $SU(5)$ grand unified theory.
In a forthcoming paper \cite{bnt} we plan to present a  more
detailed analysis of the various observations made here.

\subsection* {Acknowledgments}

\hs{-0.1cm} We thank B. Gavela, B. Grossmann, H. Frisch, A. Khanov, A.V. Kotwal, Z. Murdock,
R. Stoian and D. Zeppenfeld for helpful discussions. This work is
supported in part by US Department of Energy, Grant Numbers
DE-FG02-04ER41306 and DE-FG02-ER46140. Z.T. is also partially
supported by GNSF grant 07\_462\_4-270.

\bibliographystyle{unsrt}

\end{document}